# Title

Low-cost, portable, easy-to-use kiosks to facilitate home-cage testing of non-human primates during vision-based behavioral tasks.

# Authors

Hamidreza Ramezanpour[1,2], Christopher Giverin[1,2], and Kohitij Kar [1, 2*]

# Affiliation

1. Department of Biology, Centre for Vision Research, York University Toronto, Ontario, Canada
2. Vision: Science to Applications (VISTA), York University, Toronto, Ontario, Canada

*Correspondence should be addressed to Kohitij Kar.
 E-mail: k0h1t1j@yorku.ca

Number of Figures: 8

# Conflict of interests

The authors declare no competing financial interests.

# Acknowledgments

We thank Xiaogang Yan for assistance with cage compatibility assessments. We thank Dr. Melissa Madden, Natasha Down, and Hongying Yang for their help with NHP care and related procedures. This work has been motivated by earlier work by Elias Issa, Michael Lee, and other members of the DiCarlo Lab at the McGovern Institute for Brain Research at MIT. The work has been supported by funds from Canada Foundation for Innovation (CFI), Canada Research Chair Program, Simons Foundation Autism Research Initiative (SFARI), and the Canada First Research Excellence Funds (VISTA Program).

# New and Noteworthy

Training nonhuman primates (NHPs) for vision-based behavioral tasks in a laboratory setting is a time-consuming process and comes with many limitations. To overcome these challenges, we have developed an affordable, open-source, wireless, touchscreen training system that can be placed in the NHPs' housing environment. This system enables NHPs to work at their own pace. It provides a platform to implement continuous behavioral training protocols without major experimenter intervention and eliminates the need for other standard practices like NHP chair training, collar placement, and head restraints. Hence these kiosks ultimately contribute to animal welfare and therefore better-quality neuroscience in the long run. In addition, NHPs quickly learn complex behavioral tasks using this system, making it a promising tool for wireless electrophysiological research in naturalistic, unrestricted environments to probe the relation between brain and behavior.

# Abstract

Non-human primates (NHPs), especially rhesus macaques, have played a significant role in our current understanding of the neural computations underlying human vision. Apart from the established homologies in the visual brain areas between these two species, and our extended abilities to probe detailed neural mechanisms in monkeys at multiple scales, one major factor that makes NHPs an extremely appealing animal model of human-vision is their ability to perform human-like visual behavior. Traditionally, such behavioral studies have been conducted in controlled laboratory settings. Such in-lab studies offer the experimenter a tight control over many experimental variables like overall luminance, eye movements (via eye tracking), auditory interference etc. However, there are several constraints related to such experiments. These include, 1) limited total experimental time, 2) requirement of dedicated human experimenters for the NHPs, 3) requirement of additional lab-space for the experiments, 4) NHPs often need to undergo invasive surgeries for a head-post implant, 5) additional time and training required for chairing and head restraints of monkeys. To overcome these limitations, many laboratories are now adapting home-cage behavioral training and testing of NHPs. Home-cage behavioral testing enables the administering of many vision-based behavioral tasks simultaneously across multiple monkeys with much reduced human personnel requirements, no NHP head restraint, and provide NHPs access to the experiments without specific time constraints. To enable more open-source development of this technology, here we provide the details of operating and building a portable, easy-to-use kiosk for conducting home-cage vision-based behavioral tasks in NHPs.

# Introduction

The past few decades have seen significant advancements in systems neuroscience research, which have broadened our understanding of the brain mechanisms underlying cognitive, social, and emotional processes (Hung et al. 2005; Kar and DiCarlo 2023; Shepherd and Freiwald 2018; Tsao et al. 2008; Wang et al. 2017). Animal models, particularly non-human primates such as rhesus monkeys (Horwitz 2015; Picaud et al. 2019; Rajalingham et al. 2018), are crucial to this endeavor, given their close phylogenetic relationship to humans (Perelman et al. 2011) and the limitations involved in invasive human experimentation. However, traditional training methods for these animal subjects can be labor-intensive, time-consuming, and subject to experimenter biases, thereby introducing potential inconsistencies in research outcomes. Recent technological advancements (e.g., computationally sophisticated handheld devices with reliable connectivity via Bluetooth and Wi-Fi) offer opportunities to revolutionize this landscape through standardized, automated training platforms. These systems, often referred to as "kiosks," are set up within the animal's housing environment and provide a hands-off approach to training while also allowing for a more nuanced collection of behavioral data. Moreover, automated training in the home environment is hypothesized to reduce stress associated with transport and handling, which in turn may lead to more reliable data collection.

Prior models for automated cognitive evaluation systems designed for animal enclosures serve as references for crafting a specialized cognitive testing setup suitable for these enclosures (Griggs et al. 2021; Kell et al. 2023; Womelsdorf et al. 2021). While these prior tools have been effective in yielding significant behavioral data, many of them pose challenges in terms of practical integration within animal housing areas. For instance, they often require a separate area for computer management or are not easily removable for cleaning processes. Many of these systems are not yet publicly available, and therefore offer limited scope for incorporating new experimental frameworks, and their specialized software is often restricted to specific platforms, lacking a unified framework for assessing multiple cognitive aspects. These prior solutions often required proximity to a standard power line or the use of extension cables for operation, adding to their impracticality. Moreover, they sometimes lacked options to connect to internet or Bluetooth for remote control. To address these limitations, here we introduce a more open-sourced, touchscreen-equipped kiosk system, designed specifically for non-human primates (NHPs). One of the key advantages of the kiosk presented here is its weight, portability, and use of publicly available touchscreens such as iPads, allowing for wireless internet connectivity and battery-powered operation, significantly lowering its cost, and improving upon the limitations of other existing systems.

In this article, we summarize the operation and building blocks of the kiosk developed in our laboratory at York University (Toronto, Canada). The primary goal of this article is to make the details of our version of the kiosks readily available across the NHP (both macaque and marmoset) laboratories worldwide. An elaborate pictorial depiction of the kiosk (as provided here) and the operation videos (see Data and Code Availability) will allow laboratories to get a gist of the mechanisms involved and quickly reproduce a customized version for use. We expect that this will enhance NHP research and help fine tune our design to incorporate additional capabilities into this system.

# Material and Methods

## Subjects

### Non-human primates

Two male rhesus monkeys (*Macaca mulatta*, age: ~10 years) had 200-300 min of daily individual access to the kiosk. All data were collected, and animal procedures were performed, in accordance with the NIH guidelines, the Massachusetts Institute of Technology Committee on Animal Care, and the guidelines of the Canadian Council on Animal Care on the use of laboratory animals and were also approved by the York University Animal Care Committee.

### Human participants

We measured human behavior (88 subjects) using the online Amazon Mechanical Turk (MTurk) platform, which enables efficient collection of large-scale psychophysical data from crowd-sourced human intelligence tasks. The reliability of the online MTurk platform has previously been validated by comparing results obtained from online and in-lab psychophysical experiments (Majaj et al. 2015; Rajalingham et al. 2015). Data from these experiments have been previously published (Kar et al. 2019). All human studies were done in accordance with the Massachusetts Institute of Technology Committee on the Use of Humans as Experimental Subjects and the regulations of the ethics review board at York University.

## Interaction of the human experimenter with the kiosk

The kiosk is approximately 25-30 lbs. in weight. The National Institute for Occupational Safety and Health (NIOSH) has created a mathematical model that can predict the chance of injury when lifting weights, taking into consideration factors such as weight, frequency of lifting, and movement during lifting. A lifting equation (see http://www.cdc.gov/niosh/docs/94-110/), calculates a recommended weight limit for one individual in various situations. The lifting equation establishes a maximum load of 51 pounds, which is then adjusted based on various factors, including how often the weight is lifted, the twisting of the back during lifting, the vertical distance the load is lifted, the distance of the load from the body, the distance moved while lifting the load, and the ease of holding onto the load. Given that the kiosk can be carried around in a cart, and only needs to be lifted right before placing it on the cage bars, the overall time of lifting as well as its total weight fits adequately with the load bearing capacity of most experimenters and are within the NIOSH estimated limits. The experimenter should lift the kiosk by its side using the protruding handles, specifically added for that purpose (**Figure 1**). In an alternate model of the kiosk, the handles can be also placed vertically (instead of horizontally).

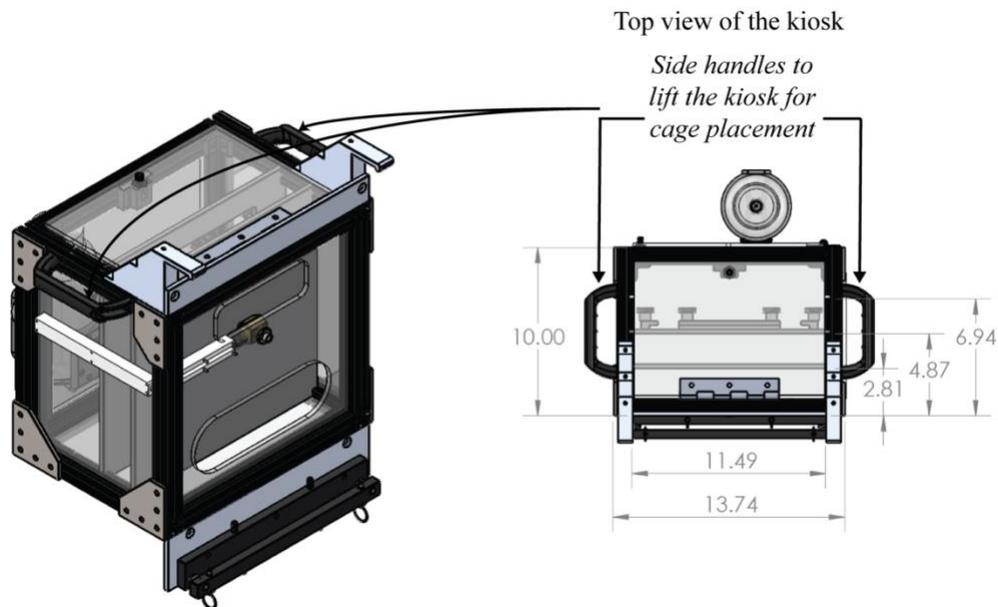

**Figure 1 Lifting the kiosk.** Primarily, the kiosk is lifted by the experimenter immediately before placing it on the cage. There are two handles on the side that allow the experimenter to lift the 25-30 lbs. kiosk.

## Software control

The kiosk can be integrated flexibly with any behavioral control software that registers touchscreen (e.g., android tablets, Apple iPads) interactions and controls the reward delivery to the animal. Here, we have used MWorks (https://mworks.github.io/), which is an open-source collection of software solutions aimed at facilitating the execution of real-time studies, notably in the fields of psychology and neurophysiology. The platform offers advanced tools that allow for the detailed planning of experimental procedures, including organizational elements like blocks and trials as well as elements of a finite state machine, such as states and transitions. It also gives researchers precise control over various types of input and output hardware, including visual displays, data collection systems, and general-purpose I/O devices. Moreover, it comes equipped with a diverse array of visual stimuli, particularly beneficial for those focusing on vision research.

## Attaching the kiosk to the cage

The kiosk needs to be securely attached (**Figure 2**A) to the top (**Figure 2**B: top panel) or the bottom (**Figure 2**B: bottom) of the cage opening with specific mechanisms (see our demo videos on GitHub, Placement.mp4 for details and part list). Please refer to the supplemental video to observe the exact locking mechanisms that prevent the kiosk from being detached from the cage (regardless of any reasonable force applied by the NHP). Given that NHP cages come in various configurations, the exact attachment styles will likely need to be modified to cater to individual caging systems.

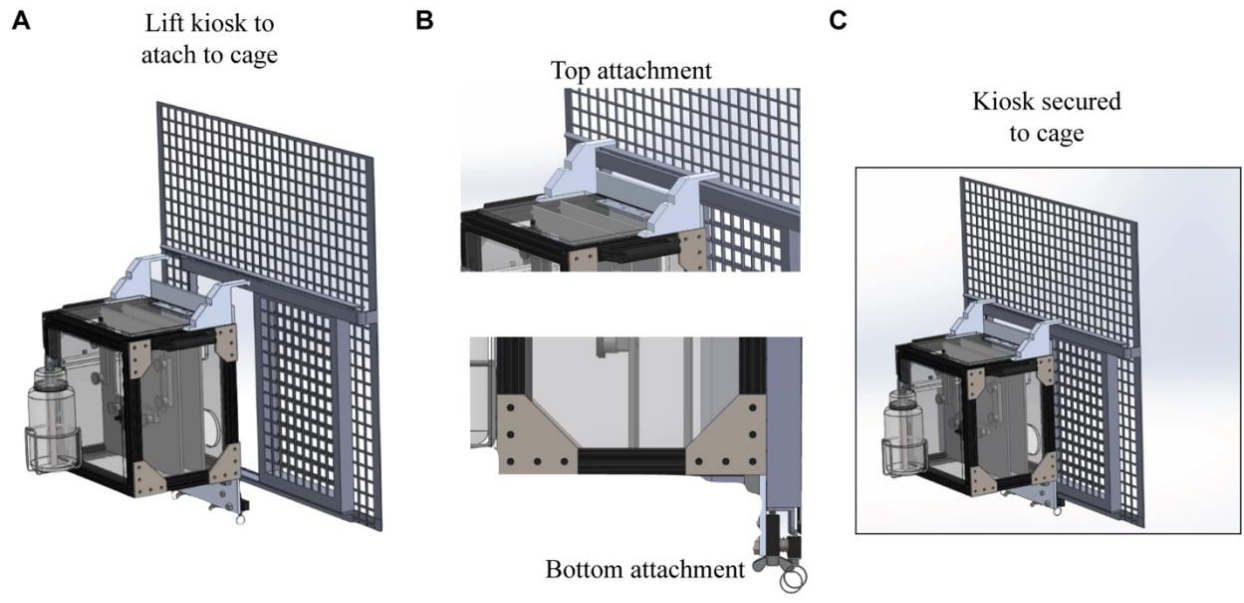

**Figure 2. Securely attaching the kiosk to the cage. A.** The kiosk can be attached to the cage **B.** The top part of the kiosk slides into the cage, while the bottom has a locking mechanism that holds it in place such that the NHP cannot move the kiosk. **C.** Once placed correctly, there are no exposed gaps that can encourage the NHP to shake the kiosk. To get a closer look at how the kiosk should be placed, kindly refer to the following video, https://github.com/vital-kolab/nhp-turk/blob/main/videos/Placement.mp4

# The internal components of the kiosk

When the monkey has access to the kiosk, that is when the kiosk is securely attached to the cage (**Figure 2C**), experimenters can still access parts of the kiosk to make necessary changes to the experiments, change out the reward fluid, replace batteries for the tablet (or other electronics, e.g., Arduinos, Raspberry Pi). Majority of the components like the side walls of the kiosk, the inserter panels, and the top and the back panel doors are all polycarbonate sheets. All parts have been itemized (see e.g., in **Figure 3**) and shared at Bill of materials.xlsx. The following are the most relevant components of the kiosks with respect to the experimenter's needs:

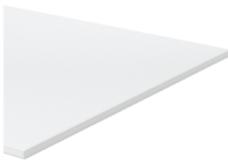

**Figure 3. Example contents of the Bill of materials**. Please find an excel sheet with detailed information on all of these items for each part at https://github.com/vitalkolab/nhp-turk/blob/main/Bill.of.Materials.xlsx

## The top panel (non-detachable from the kiosk)

The top panel door can be pulled open to access the content of the kiosk (**Figure 4**A). Typically, one would open it in the beginning of the experimental session to insert the tablet/touch screen into the kiosk (**Figure 4**B). The experimenter can also place dry treats and toys into the front end of the kiosk during initial days of the NHP training to make the NHP feel motivated to insert their hands into the kiosk. They can also adjust the electronics at the back of the kiosk.

## The back panel (non-detachable from the kiosk)

The back panel allows the experimenter to insert the electronics (**Figure 4**E) associated with running the experiments. It is also attached to the section that contains the reward fluid bottle (**Figure 4**D). The bottle has a tube inserted in it that is usually connected to a pump and then via another tube to a front nozzle (that the NHP uses to get the reward).

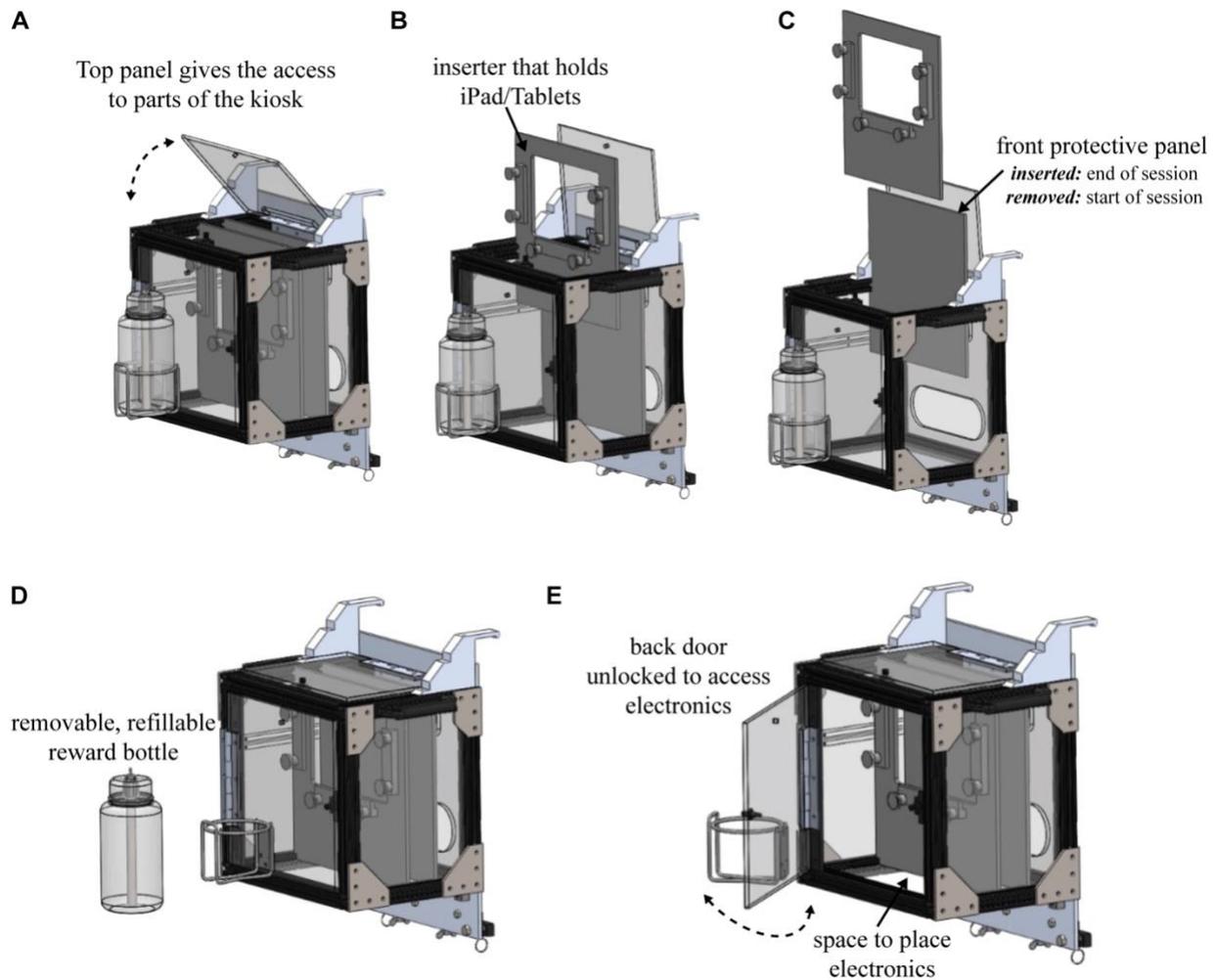

**Figure 4. Interacting with the kiosk. A.** The top panel that remains attached to the kiosk can be opened to access the rest of the kiosk. **B.** One of the detachable inserter panels holds the tablet/touch screen. **C.** A protective front panel is inserted at the end of the experimental session to protect the electronics at the back of the kiosk from the monkey. It is also removed at the beginning of the session to allow the monkey full access to the touch screen. **D.** A removable bottle is placed in a dedicated attachment in the back of the kiosk to hold the reward fluid (typically water). The bottle can be removed daily to refill the fluid. **E.** The back door (that stays attached to the kiosk) can be opened to insert the electronics, and make proper adjustments (e.g., tighten the tablet into the inserter).

## The two inserter panels (removable from the kiosk)

The kiosk uses two panels that can be inserted into it. The front panel (**Figure 4**C) is for protective purposes, and it is only inserted at the end of the experimental session, so that the monkey cannot access the electronics at the back of the kiosk. Typically, the tablet is removed for charging at the end of the day. Therefore, the front panel has to be placed. This panel is also removed at the beginning of the experimental session so that the NHP can access the touch screen. The second inserter panel (**Figure 4**B) holds the tablet/touch screen. This panel ideally could be left in the kiosk at all times. The back of this panel (that is accessible to the experimenter when the NHP has access to the kiosks), has brackets and knobs to securely place an iPad or other tablets into that slot (see **Figure 5**A; left panel). The front of this panel has an aperture sized according to the dimensions of the touch screen (**Figure 5**A; right panel). Depending on whether the vision tasks are conducted on iPads or Android tablets or via other means, the size of the aperture should be adjusted. The current dimensions of this inserter panel are provided in **Figure 5**C. Other features to the kiosks, like eye tracking can be added by making modifications to this inserter panel. For instance, an example strategy could involve making three additional holes on this inserter panel. Two of these could be at the bottom to accommodate two IR illuminators and one could be at the top to allow for a HD USB-based camera with IR filters (**Figure 5**B). A combination of these can be used to set up a low-cost eye tracker (more details of this tracker will be available soon).

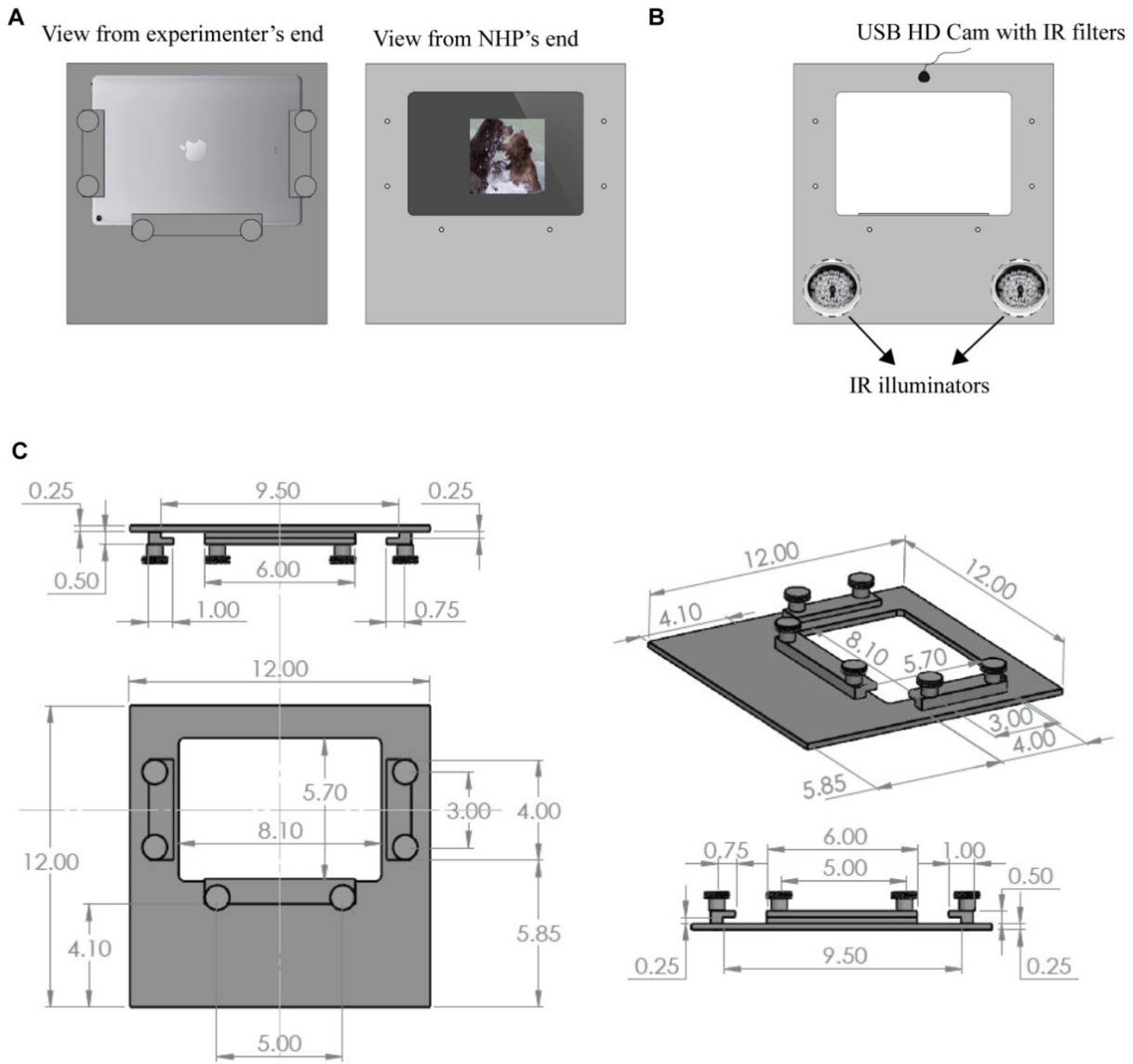

**Figure 5 . The touch screen/tablet inserter panel. A.** Left Panel: The view from the experimenter's end. The knobs can be loosened to insert the tablet from the top of the kiosk. The back of an iPad is visible in the figure as an example configuration. Right Panel: The view from the monkey's end. The visual stimulus

can be seen in the example. The monkeys can touch this screen to indicate their choices. **B.** Possible items to add to the setup via this inserter. Apertures can be drilled at the bottom to allow for two IR illuminators. A small hole at the top of the inserter panel can be used to insert a high-definition USB camera with an IR filter. These can be combined to develop a low-cost eye tracking system (available in future versions of the kiosk). **C.** The dimensions of the touch screen/tablet inserter panel. The exact dimensions can vary based on the make of the tablet used. The aperture size can be smaller than the touch screen if the entire screen is not utilized in the tasks. However, the size shouldn't be larger than the touch screen since that will allow the monkeys to access the electronics (usually placed behind the touch screen).

## The reward bottle attachment

A removable, refillable reward (water, juice etc.) bottle can be attached to the back of the kiosk (**Figure 4D**). It is advised that the bottle be kept empty while transferring the kiosk for cage attachment and only filled up once it is securely attached. This will significantly reduce the overall weight of the kiosk system.

## Interaction of the animal with the kiosk

To begin the experimental session, the kiosk must be first securely attached to the cage (**Figure 6**A). The kiosk can be placed both at the top-half or the bottom half of a standard NHP caging system. If it is placed in the top-half, a horizontal divider should be in place to allow the monkey to rest on (**Figure 6**B) while they perform the behavioral tasks. The animals have access (**Figure 6**B) to a touch screen (e.g., android tablets, Apple iPads) via two apertures (**Figure 6**C) in the front of the kiosk. The top aperture (see **Figure 6**C for dimensions) provides a clear, non-occluded visual access to the screen. The bottom aperture allows the monkey to reach inside with their hands and use the touch screen. Of note, it is important to make sure that the edges of this aperture have been additionally smoothened by the use of sandpaper (or something similar), since rubbing along those edges (if not polished) might lead to minor cuts in the NHP arms. It is usually advised that the experimenter tests the kiosk in a lab setting with themselves (as subjects) first to ensure that everything is working correctly. In the beginning of the NHP training pipeline, the space between the touchscreen and the front screen may be utilized to place toys and treats (**Figure 6**D) so that the monkey feels motivated (and not afraid) to insert their hands through the aperture

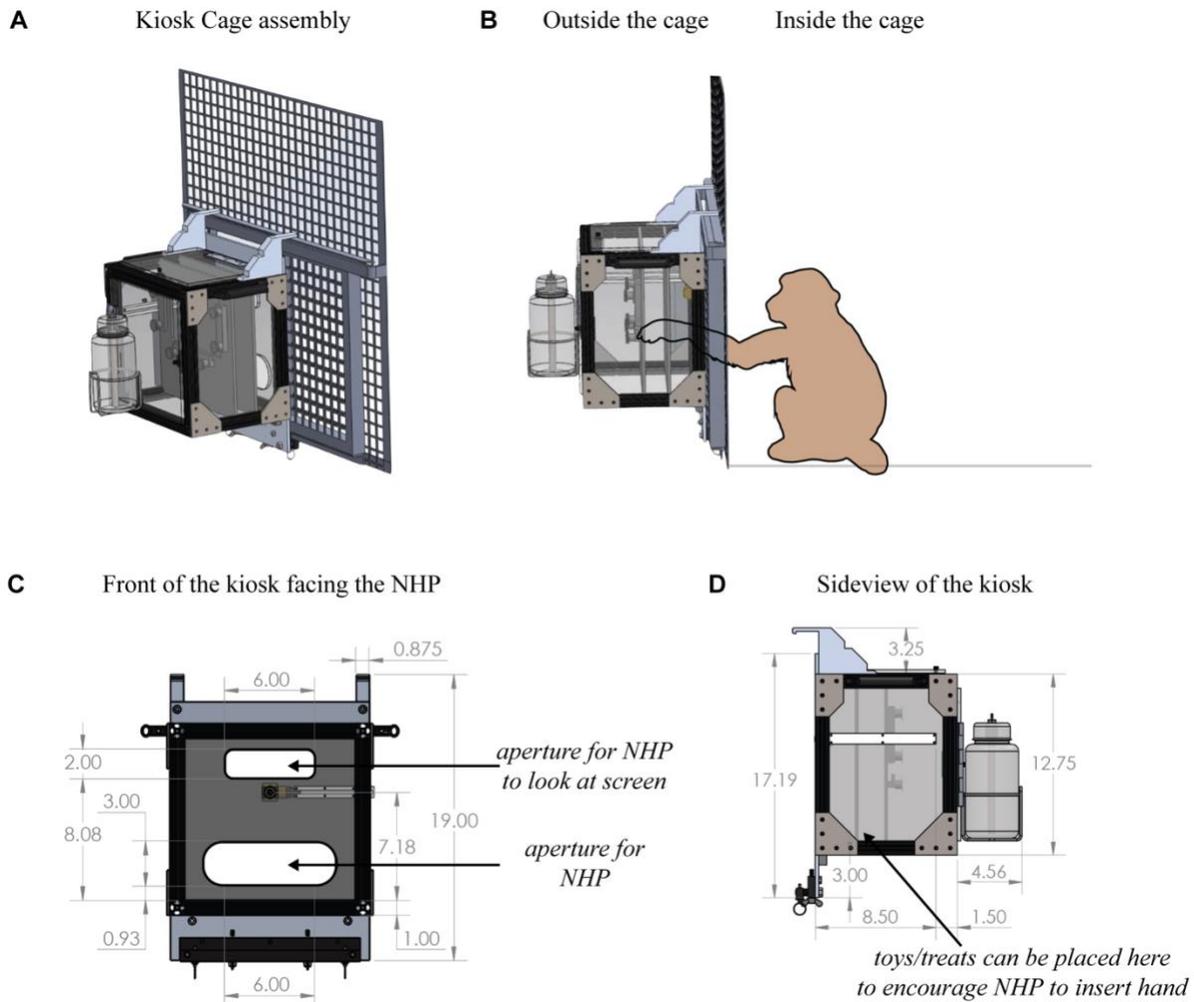

**Figure 6. Kiosk components critical for NHP interactions. A.** Kiosk Cage assembly. Before the NHP is allowed to interact with the tablet by removing the front-most protective inserter panel, the kiosk has to be securely attached to the cage **B.** The monkey accesses the touch screen by sitting right in front of the kiosk on a horizontal cage divider panel. **C.** The front of the kiosk facing the NHP has two apertures. The top one allows for clear visual access to the NHP. The bottom one allows the NHP to insert its hand and interact with the touch screen. **D.** Sideview of the kiosk illustrates the space at the front that can be used to insert toys and treats to initially encourage the NHP to get their hands in via the aperture.

## Behavioral Tasks

In this study, we used a set of 800 naturalistic ('synthetic') images. Each image consisted of a two-dimensional (2D) projection of a three-dimensional (3D) model (purchased from Dosch Design and TurboSquid) added to a random background. The ten objects chosen were bear, elephant, face, apple, car, dog, chair, plane, bird, and zebra. By varying six viewing parameters, we explored three types of identity while preserving object variation, position (x and y), rotation (x, y, and z), and size. All images were achromatic with a native resolution of $256 \times 256$ pixels. A total of 800 naturalistic images (80 per object category) were used.

## NHP in Lab

We measured monkey behavior from two male rhesus macaques. Images were presented on a 24-inch LCD monitor (1,920 × 1,080 at 60 Hz) positioned 42.5 cm in front of the animal. Monkeys were head fixed. Monkeys fixated on a white circle (0.2°) for 300 ms to initiate a trial. The trial started with the presentation of a sample image (from a set of 800 images) for 100 ms. This was followed by a blank gray screen for 100 ms, after which the choice screen was shown containing a standard image of the target object (the correct choice) and a standard image of the distractor object. The monkey was allowed to freely view the choice objects for up to 1,500 ms and indicated its final choice by holding fixation over the selected object for 400 ms. Trials were aborted if the gaze (monitored by an Eye Link 1000, eye tracker system) was not held within ±2° of the central fixation circle during any point until the choice screen was shown.

## NHP in home-cage

The home cage behavioral tasks were very similar to the in-lab tasks. There were two main differences. First, we did not track the eyes of the monkeys in the cages. Second, the monkeys indicated their choices by using the touch screen and tapping on the choice object images (similar to the human participants on MTurk).

## Humans on MTurk

Each trial started with a 100-ms presentation of the sample image (1 out of 800 images). This was followed by a blank gray screen for 100 ms followed by a choice screen with the target and distractor objects, similar to a previous study (Kar et al. 2019). The subjects indicated their choice by touching the screen or clicking the mouse over the target object. Each subject saw an image only once. We collected the data such that there were 80 unique subject responses per image with varied distractor objects.

## Behavioral Metrics

We have used a one-vs-all image level behavioral performance metric, $B.I_1$, similar to a previous study (Rajalingham et al. 2018), to quantify the behavioral performance of the monkeys and humans (described below). This metric estimates the overall object discriminability of each image containing a specific target object from all other objects (pooling across all 9 possible distractor choices).

Given an image of object '$i$', and all nine distractor objects ($j \neq i$) we computed the average performance per image as the average of the percent correct across all the binary tasks done with that image as the sample image (where object 'i' was the target and all objects $j \neq i$ were the distractors respectively).

To compute the reliability of this vector, we split the trials per image into two equal halves by resampling without substitution. The median of the Spearman-Brown corrected correlation of the two corresponding vectors (one from each split half), across 100 repetitions of the resampling was then used as the reliability score (i.e., internal consistency). The ceiling (bold horizontal lines in **Figure 8B and C**) for any correlation was computed as the squared root of the product of the reliability of the two variables involved in the correlation.

# Results

In the section below, we demonstrate the similarity in the behavioral results obtained across similar experiments conducted in macaques across different environmental settings (in the laboratory rigs and in the home-cages). We also assess how human behavior compares to the macaques across these settings. The overlapping behavioral variances observed across these two species (highlighted in red; middle panel; **Figure 7**) may limit the extent to which neural mechanisms, deduced through invasive studies in the macaque, are applicable in explaining human behavior.

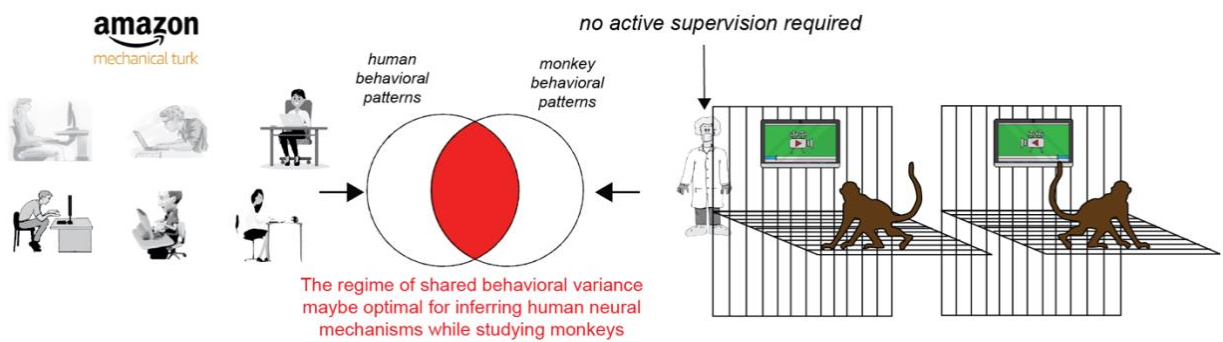

**Figure 7. Comparison of behavioral benchmarks across humans and monkeys can provide guidance on how to optimally use rhesus macaques as a model of human cognition.** We measured behavioral accuracies on an object discrimination task on MTurk (left panel) across 80 human participants. We measured monkey behavior (n=2) on the same task in their home cages (right panel). Unlike measurements done within the laboratory rigs, this method significantly reduces the requirement of research personnel to actively monitor the animals. The overlap in the behavioral variances measured across these two species (indicated in red; mid panel) might constrain how well neural mechanisms inferred via invasive studies in the macaque is relevant to explaining the human behavior.

## Reliability of macaque behavioral measurements in-cage and in-lab

To assess the quality of measurements in the lab and in the home cages, we computed the trial split half reliability of the $B.I_1$ estimates in these two settings (blue: in-lab, red: in-cage) as a function of total number of trials. As expected, in-lab studies (**Figure 8A**) produce a small but significant edge over in-cage measurements. These differences are likely due to more controlled settings in the lab. However, as evident in the scaling trends, the in-cage data reliability increases with more repetitions of the trials.

## Comparison of in-cage and in-lab behavior of macaques

To validate our methods, we first compared the behavior measured in the home cage and in the more controlled setting of the lab in two monkeys. **Figure 8B** shows that monkey 1 and monkey 2 have significantly high correlation (monkey 1: Pearson R = 0.61, ceiling = 0.59 ± 0.04; monkey 2: Pearson R = 0.60, ceiling = 0.71±0.02) for the image-level object discrimination accuracies ($B.I_1$). These results demonstrate the significant similarity in the behavior measured across these two settings where the method of reporting the choices also varies (in cage: touch screen, in lab: eye fixations).

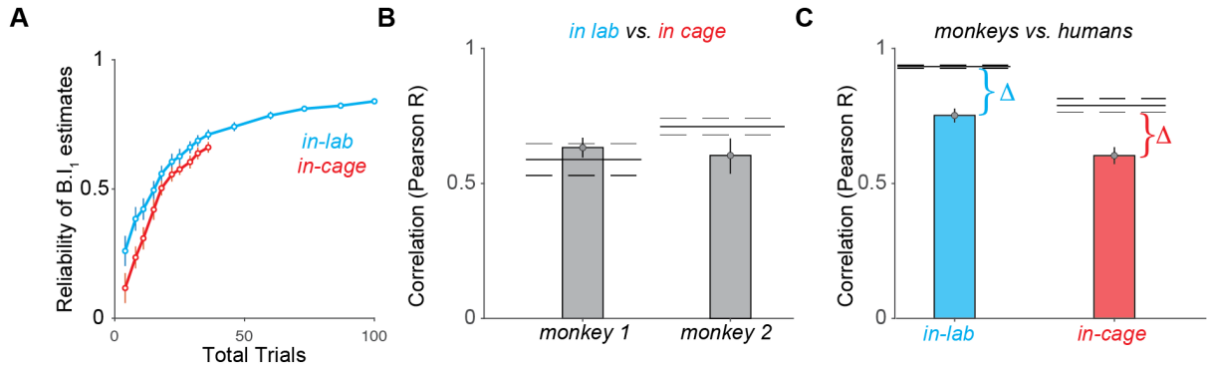

**Figure 8. Comparison of behaviors measured in the lab, in the monkey home cages and in humans over Amazon Mechanical Turk. A.** Reliability of macaque behavioral measurements across different settings. Trial split half reliability of the $B.I_1$ estimates are shown as a function of the total number of trials for both in-lab (blue) and in-cage (red) settings. In-lab measurements present a slight advantage over in-cage settings, likely attributed to the controlled environment of the lab. However, the in-cage data reliability demonstrates increasing consistency with a higher number of trial repetitions. **B. Intra-species comparison of macaque behavior in different environments.** Behavioral measurements for monkey 1 and monkey 2 across the in-cage and in-lab settings are presented. Notably, both monkeys exhibit high correlation in their image-level object discrimination accuracies ($B.I1$) irrespective of the environment (monkey 1: Pearson R = 0.61, ceiling = 0.59 ± 0.04; monkey 2: Pearson R = 0.60, ceiling = 0.71±0.02). These results underscore the considerable similarity in behaviors, even with varied reporting methods, such as touch screens in cages versus eye fixations in the lab. **C. Inter-species comparison between human and macaque behaviors.** A direct comparison of $B.I_1$ estimates across human participants on Amazon Mechanical Turk (MTurk) and monkeys in both settings (in-cage and in-lab) was conducted. The observed differences between human and monkey behaviors remain statistically insignificant across the two measurement environments, with values of Δ (human vs. monkeys: in-lab) = 0.1814 and Δ (human vs monkey: in-cage) = 0.1859. The error bars represent standard deviation of correlations (bootstrapped across images), and horizontal dashed lines represent standard deviation of ceiling.

## Comparison of human with monkey behavior measured in the lab and in their home cages.

Next, to test whether the relationship between behavior measured across human participants (on MTurk) and monkeys (in-cage and in-lab) are consistent across in-lab and in-cage measurements we directly compared the $B.I_1$ estimates across these conditions. We observed that the difference between humans and monkey behavior was not significantly different when measured under these two environments (**Figure 8**C; Δ (human vs. monkeys: in-lab) = 0.1814, Δ (human vs monkey: in-cage) = 0.1859).

# Discussion

In this study, we have successfully introduced and provided a thorough implementation guide for a low-cost, portable kiosk designed for home-cage vision-based behavioral tasks in rhesus macaques. Traditional methodologies for studying NHP behavior have several limitations including the need for dedicated experimenters, invasive surgeries, and limited experimental time, among others. As we have demonstrated, our home-cage behavioral testing system significantly mitigates these issues while enabling the collection of behavioral data that is well aligned with the results obtained in more controlled studies in the laboratories. Our kiosk addresses several of these limitations, offering a streamlined, efficient approach to visual behavior research without the complications associated with laboratory settings. With iPads (or any other similar tablets) for its touchscreen interface, the kiosk provides an unprecedented level of portability, flexibility, and technological prowess that makes it easy to integrate into varied research environments. Open-source software compatibility not only ensures the kiosk's accessibility to labs worldwide but also fosters a collaborative atmosphere where researchers can share advancements and modifications. We believe that the open-access nature of our design will motivate further innovation in NHP research settings, which will continually refine and expand the capabilities of such systems.

Furthermore, the well-documented and elaborately depicted construction and operation details are aimed at helping laboratories (across the world) with access to mechanical workshops to easily replicate the setup. By doing so, we aim to make these kiosks a staple in NHP research, thereby reducing the cost, time, and ethical concerns traditionally associated with primate cognitive studies.

A distinct advantage of the rhesus macaque animal model is its compatibility with human behavioral studies, as demonstrated in our results and in prior work (Kar et al. 2019; Kar and DiCarlo 2021; Rajalingham et al. 2018). This specifically makes macaques an appropriate animal model for neurological disorders like autism spectrum disorder (ASD) where diagnosis is primarily based on human behavioral markers. As preclinical genetic ASD models of macaques are being developed, the ability to test them readily on appropriate human-compatible behavioral paradigms will likely be extremely relevant. Our system enables to implement such protocols at scale. In addition, in previous work (Kar 2022), we have also elaborated on how artificial neural network models of vision can be leveraged to optimize these efforts. While the current design allows for detailed behavioral studies (measuring reaction times, choices etc.), the integration of eye tracking would further enhance the quality of data by providing insight into gaze patterns, especially relevant for autism research (Nakano et al. 2010), and attentional processes in the NHPs (Ramezanpour and Fallah 2022; Ramezanpour and Thier 2020). The touchscreens used, such as iPads, already possess the computational power to run basic eye-tracking algorithms, and this could be a significant addition to the system's capabilities. Second, the implementation of a face recognition system or an intradermally placed RF-chip based tagging of animals could facilitate multi-animal usage within the same home-cage. Such a feature would allow individual animals to log in and engage with the kiosk independently, enabling researchers to collect data from multiple subjects simultaneously without the need for manual identification or segregation.

Another noteworthy feature of the kiosk system is its connectivity capabilities. Given that the tablet, such as an iPad, can be connected to the internet via Wi-Fi, researchers can remotely monitor the status of ongoing experiments in real-time. This eliminates the need for constant physical checks on the kiosk and allows for more efficient use of personnel resources. Additionally, this connectivity enables real-time data collection and analysis, providing researchers with the flexibility to make timely adjustments to the experimental design if needed. In essence, the Wi-Fi-enabled connectivity not only simplifies logistical considerations but also opens the door for dynamic, adaptive research protocols.

The kiosk presented here should be considered as work in progress. As we develop the system further, this article will be updated with that information. We encourage the NHP neuroscience research community to contact us with suggestions or clarification questions. This will help us improve the system and make it more versatile. Our next goals are to add a low cost eye tracking system, or even adapt webcam based alternatives (Papoutsaki 2015), as mentioned in **Figure 5B**, and to release the codebase and material for the electronics that we are using. Our kiosks are fully compatible with behavioral testing of macaques paired with chemogenetic perturbations (Kar et al. 2021), and wireless electrophysiological recordings. As we further develop those studies in our laboratory, we shall update the system details.

# Data and Code Availability

The details of the kiosk can be found at the following GitHub page: https://github.com/vital-kolab/nhp-turk